\documentclass[aps,prl,groupedaddress,fleqn,twocolumn,longbibliography]{revtex4}
\pdfoutput=1
\usepackage{graphicx}
\usepackage{xcolor}
\usepackage{ulem}
\usepackage{amsmath}
\usepackage{gensymb}
\usepackage{float}
\usepackage{physics}
\usepackage{amsfonts}
\newcommand{\comment}[1]{}

\begin{document}
\title{Fast dynamics and high effective dimensionality of liquid fluidity}
\author{C. Cockrell$^{1,*}$, O. Dicks$^{2}$, I. T. Todorov$^{3}$, A. M. Elena$^{3}$ and K. Trachenko$^4$}
\address{$^1$ Department of Materials, Imperial College London, Exhibition Road, London, SW7 2AZ, UK}
\address{$^2$ Department of Physics and Astronomy, University of British Columbia, Agricultural Road, Vancouver, V6T 1Z1, Canada}
\address{$^3$ Daresbury Laboratory, Keckwick Ln, Daresbury, Warrington, WA4 4AD, UK}
\address{$^4$ School of Physical and Chemical Sciences, Queen Mary University of London, Mile End Road, London, E1 4NS, UK}


\begin{abstract}
Fluidity, the ability of liquids to flow, is the key property distinguishing liquids from solids. This fluidity is set by the mobile transit atoms moving from one quasi-equilibrium point to the next. The nature of this transit motion is unknown. Here, we show that flow-enabling transits form a dynamically distinct sub-ensemble where atoms move on average faster than the overall system, with a manifestly non-Maxwellian velocity distribution. This is in contrast to solids and gases where no distinction of different ensembles can be made and where the distribution is always Maxwellian. The non-Maxwellian distribution is described by an exponent $\alpha$ corresponding to high dimensionality of space. This is generally similar to extra synthetic dimensions in topological quantum matter, albeit higher dimensionality in liquids is not integer but is fractional. The dimensionality is close to 4 at melting and exceeds 4 at high temperature. $\alpha$ has a maximum as a function of temperature and pressure in liquid and supercritical states, returning to its Maxwell value in the solid and gas states.
\end{abstract}

\maketitle


\section{Introduction}

Similarly to gases and differently from solids, liquids flow. However, important liquid properties such as density, bulk modulus and heat capacity, especially at conditions not far from melting, are very close to those of solids but are very different from those of gases. This dichotomy has contributed to the problem of theoretical description of liquids and the long-held belief that a general liquid theory is unworkable \cite{landaustat} and not tractable due to the absence of a small parameter \cite{Pitaevskii,akhiezer,ropp}. This is markedly different from solids and gases where a general theory, universally describing their properties, was developed long ago.

The dynamics of individual liquid particles also shows a dichotomy related to solid and gas dynamics first introduced by Frenkel \cite{frenkel}: the motion of liquid particles combines small solidlike oscillations around quasi-equilibrium positions inside a cage formed by particle neighbours and larger-distance jumps to new positions. Observed in molecular dynamics simulations and referred to as ``transits'' \cite{wallacemd}, these jumps endow liquids with their ability to flow \cite{dyre}. The distance travelled by transits is close to the inter-particle separation as envisaged by Frenkel \cite{frenkel,wallacemd}.

In order to understand liquids at the level similar to that existing in solids and gases, we need to understand microscopic dynamics of liquid particles. This dynamics includes individual particle velocities and sets the system trajectory in phase space and all of system properties \cite{landaustat}. This dynamics also underpins properties averaged over different particles and time such as correlation functions often discussed in the area of liquids \cite{hansen1,balucani}. Useful in some cases, averaging in liquids, whose dynamics combine oscillation and transits, inevitably lose detailed information about these distinct components of motion contained in the primary phase trajectory data. 

Conceptually, an oscillatory motion is well-understood and is at the cornerstone of physics. While it took time to ascertain this motion in liquids, accumulating experiments show that the oscillatory motion in liquids is very similar to that in solids. Strong evidence for this includes close similarity of phonon dispersion curves measured in liquids and solids \cite{copley,pilgrim,burkel,pilgrim2,water-tran,hoso,hoso3,monaco1,monaco2,sn,ropp}. This similarity extends to the zone boundary as in solids, corresponding to particle vibrating inside its cage as envisaged by Frenkel.

Differently from the oscillatory component of motion, the microscopic nature and properties of diffusive transits in liquids remain largely unknown and are unclear even at the conceptual level. This includes the transit energy, distribution of this energy across different transits and so on. The problem is that we can't use results for flowing diffusive transits from the dense gas theory \cite{chapman}: differently from gases, liquids are condensed state of matter. This implies that transits in liquids constantly move in a strong fluctuating field created by other particles, with the inter-particle interaction being as strong as in solids. For this reason, useful concepts such as sudden collisions and related mean free path used in the theory of gases do not apply to dense strongly-interacting liquids \cite{chapman}. 

Here, we find that transits in liquids have several unusual and unexpected properties. First, they move on average faster than the the whole system. Second, the distribution of transit velocities is non-Maxwellian. Third, this distribution is characterised by an exponent $\alpha$ corresponding to effective space dimensionality larger than $3$. This is generally similar to extra synthetic dimensions in topological quantum matter \cite{naturereview,phystoday}, albeit extra dimensionality in liquids is not integer but is fractional. Interestingly, $\alpha$ has a maximum as a function of temperature and pressure, reducing to the normal Maxwell value in solid and gas states. This effect is unrelated to dynamic heterogeneity discussed in the slow supercooled regime of liquids approaching glass transition.

\section{Methods}

We perform molecular dynamics (MD) simulations using the DL$\_$POLY package \cite{dlpoly}. The interactions in our system are set by the widely-used Lennard-Jones potential, with parameters $\epsilon = 0.003177$ eV and $\sigma = 2.775$ \AA\ corresponding to neon. We equilibrate systems with periodic boundaries along isobars and isotherms in the NPT ensemble, and use the average density of these runs to generate configurations for use in production runs in the NVE ensemble. We select the NVE ensemble because we are interested in the individual dynamics of atoms, and therefore cannot allow thermostats or barostats interfering with the natural Hamiltonian evolution of the system. In all simulations a timestep of 1 fs was employed. Systems with 500 to 32000 atoms were used in this study. Results discussed below are size-independent.

The main property we will be concerned with is the distribution of particle velocities of transit atoms, $\rho(v)$, where $v=|{\bf v}|$ and ${\bf v}$ is velocity. We start with describing an algorithm to detect transits. As mentioned earlier, the main property of a transit is its large displacement (compared to a smaller oscillatory motion inside the cage): the transit moves from one quasi-equilibrium position to another, endowing the liquid with the ability to flow. This implies that transits are to be detected based on the distance travelled, $d$: $d>d_c$, where $d_c$ is the distance cutoff.

With this definition, we calculate $\rho(v)$ from the molecular dynamics trajectory, which we print every 10 fs. At the moment an atom's position satisfies the transit criterion, we log its instantaneous velocity. Because we work with fluids which lack a crystal structure, the reference structure for a transit at the start is simply the initial condition of the simulation. The 10 fs printing frequency is a compromise between accuracy and computational load. Frequent printing allows the capture of a jump as it happens, rather than after its conclusion. Test runs with more frequent printing - every timestep (1 fs) - did not yield any difference in the results which follow. We then construct a histogram from this list of transit velocities with a bin width of 1 m/s, which is normalised to create a probability distribution $\rho(v)$. Once an atom has been flagged as in-transit, its starting position could be reset within the reference structure in order to collect its velocity again for detection of subsequent transits. In practice, we improve statistics by limiting simulation time to 30 ps and running between 200 and 4000 different initial conditions (using different velocity seeds), such that the final distributions are composed of, typically, 500000 to 2000000 velocities, depending on the desired degree of noise. This gives the same distribution as a single long simulation where atomic reference positions are reset after a transit.

We note that there is no time averaging involved in detecting transits: we do not add a condition of the particle moving a certain distance within a certain timeframe. Instead, we log the instantaneous velocity at the moment the atom is designated as a transit. This is essential in order to study the dynamics of mobile transits.

\begin{figure}
{\scalebox{0.55}{\includegraphics{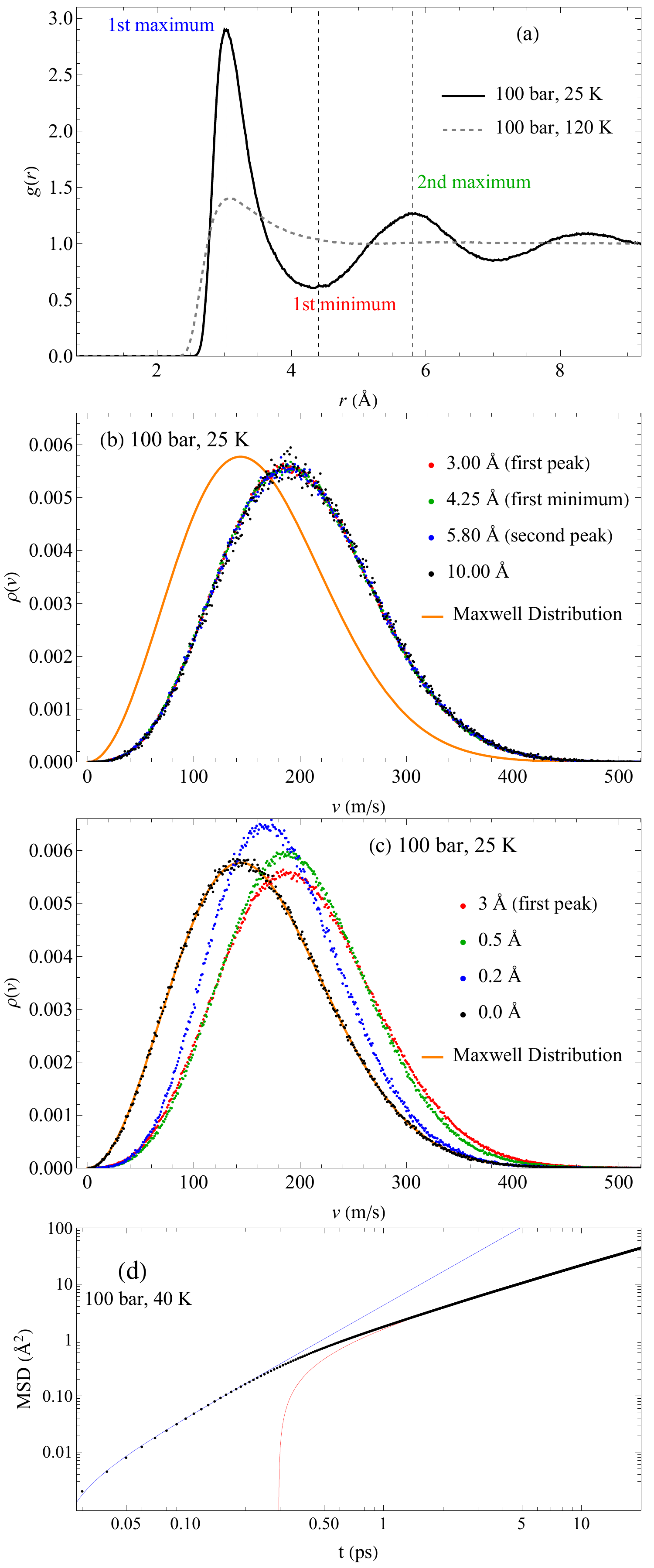}}}
\caption{(a) RDF at 100 bar in the liquidlike state at 25 K and the gaslike state at 120 K; (b) Distribution of transiting atoms with cutoffs corresponding to the maxima and minima of the RDF as well as a large cutoff of 10 \AA; (c) Distribution of transiting atoms with small cutoffs, showing convergence onto the Maxwell distribution of the whole system.}
\label{fig:cutoffs}
\end{figure}

We show the radial distribution function (RDF) at 100 bar and two different temperature in Fig. \ref{fig:cutoffs}a. The low temperature $T=25$ K marks several points, including the first maxima and minima. We use these distances as useful reference points for cutoffs $d_c$. $\rho(v)$ of transits calculated for $d_c$ in the range 3-10 \AA\ are shown in Fig. \ref{fig:cutoffs}b (using larger $d_c$ is limited by the simulation time needed to acquire long diffusion distances, especially at low temperature). We observe that the distributions are very close to each other. This was tested for other conditions explored in this study. For sufficiently large $d_c$, the algorithm catches an atom undergoing its 1st, 2nd, 3rd and other subsequent transit motions. These consecutive transits are physically equivalent to each other. Hence it doesn't matter if we record an atom's velocity during its first or any subsequent transits. This explains the insensitivity of $\rho(v)$ to $d_c$. We note that the algorithm could catch atoms in the middle of oscillatory motion between the transits, however a large-amplitude transit is much more likely to cross a large distance as compared to a small-amplitude oscillating atom. The convergence of $\rho(v)$, starting from sufficiently large $d_c$, into a single function indicates that these distributions represent a physically meaningful sub-system of atoms.  

Differently from large $d_c$, $\rho(v)$ deviate at small $d_c$, as displayed in Fig. \ref{fig:cutoffs}c, and start to visibly converge towards the Maxwell distribution of the whole system below about 1.0 \AA\ and for $d_c$ closer to amplitude of the the oscillatory motion. This is easy to understand: short $d_c$ include oscillatory motion, and therefore the motion of nearly any atom will satisfy the transit criterion. This cutoff at which the distribution of transit atoms becomes cutoff-independent, 1.0 \AA, corresponds to the approximate maximum amplitude of oscillatory motion. An atom which meets this displacement is therefore undergoing a non-oscillatory transit. This interpretation is ratified by the plot of the mean square displacement (MSD) as a function of time in Fig. \ref{fig:cutoffs}d. The nonlinear short-time behaviour is well-known to correspond to the free motion of a particle before it interacts with and is accelerated by its neighbours. This nonlinear regime therefore roughly corresponds to the amplitude of oscillation, and we fit a quadratic function to it. The long-time behaviour is the combination of oscillation and transits regime, and to it we fit a linear function. The departure from the quadratic ballistic regime visibly begins at around 0.6 \AA, and concludes shortly after 1.0 \AA, which coincides with the convergence of $\rho(v)$ to its high-$d_c$ value at around 0.8-1.0 \AA. The distribution $\rho(v)$ therefore becomes constant when the MSD switches from its ballistic to its diffusive regime.

This implies that $\rho(v)$ are insensitive to the definitions of transits for sufficiently large $d_c$ and can be consistently used in our analysis. We note a related way to characterise transit motion used earlier. These characterisations involved the calculation of the liquid relaxation time $\tau$, the time it takes the transit to move from one quasi-equilibrium point to the next \cite{frenkel}. Tuning $d_c$ gives $\tau$ similar to those based on other methods such as decay of correlation functions \cite{sastry}. Previous work \cite{egami} used a topological definition of $\tau$ as time between an atom gaining or losing a nearest neighbour. We calculated velocity distributions of atoms which satisfied this criterion, again using a variety of $d_c$. We found that such a distribution is unfailingly identical to the Maxwell distribution because the histogram $\rho(v)$ now includes oscillating atoms whose motion contributes to an atom gaining or losing a neighbour. We combined this topological definition with our geometric definition: an atom was defined as in-transit if it underwent a neighbour change \textit{and} had been displaced beyond a certain $d_c$. The displacement cutoff and the neighbour cutoff were defined to be equal to one another. This definition yielded the same distributions as those resulting from the displacement criterion alone, at all pressure and temperature conditions.

\begin{figure}
{\scalebox{0.55}{\includegraphics{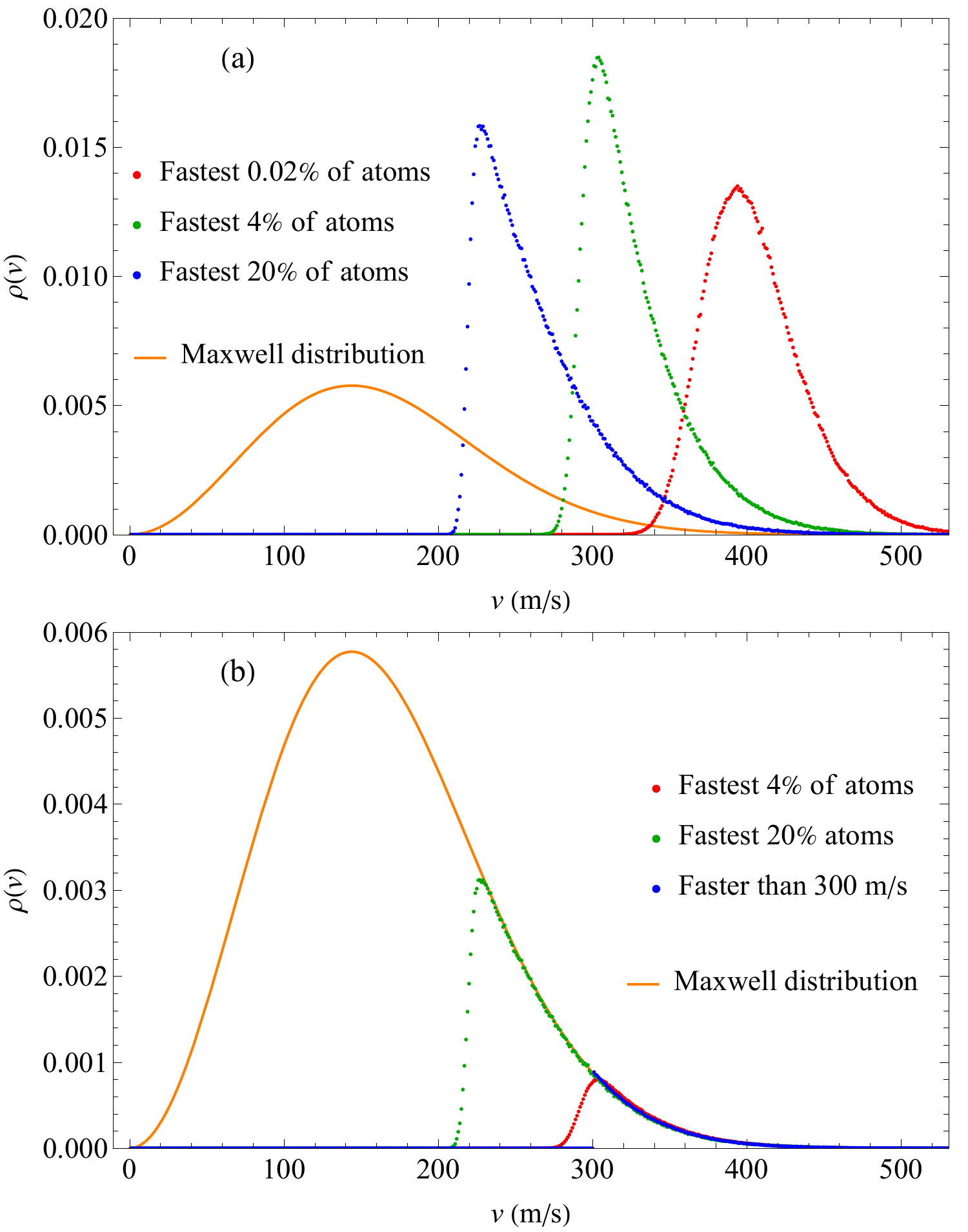}}}
\caption{(a) Probability density of the fastest 0.02\%, 4\% and 20\% atoms in the system; (b) Distributions of fastest subsets rescaled such that their convergence onto the tail of the Maxwell distribution is evident, as well as the distribution formed by explicitly selecting atoms faster than 300 m/s.}
\label{fig:velocitycriteria}
\end{figure}

Recall that our selection criterion of transits is not based on particle's velocity and therefore is not related to selecting faster atoms directly. The resulting $\rho(v)$ converge into a single cutoff-independent function, which is not the case for the criteria based on velocity. In Fig. \ref{fig:velocitycriteria}a, we plot the velocity distribution formed by logging the velocities of the fastest subsets of atoms in the system at each timestep. These distributions depend heavily on the subset size: as this size becomes large, the distributions become the tail of the Maxwell distribution, as they should. This is demonstrated in Fig. \ref{fig:velocitycriteria}b where $\rho(v)$ from Fig. \ref{fig:velocitycriteria}a are re-scaled to the tail of the Maxwell distribution. Fig. \ref{fig:velocitycriteria}b also includes a distribution formed from atoms moving faster than 300 m/s. We therefore find that velocity-based criteria do not converge on a single distribution, whereas the physically justified criterion based on large distance covered by transits do.

\section{Results and Discussion}

We now analyse the velocity distributions of transits. The central observation from Fig. \ref{fig:cutoffs}b is that $\rho(v)$ of transits is shifted to the right relative to the Maxwell distribution, implying that transits move on average faster than atoms in the overall system. We use nonlinear fitting in Mathematica with a generalised Maxwell distribution as a fitting function:

\begin{equation}
\rho(v)\propto v^\alpha \exp(-\frac{mv^2}{2k_{\rm B}T})
\label{d1}
\end{equation}

When $\alpha = 2$, Eq. \eqref{d1} is the standard Maxwell distribution in three dimensions \cite{landaustat}. If $\rho(v)$ is calculated over all atoms in the system, such as seen in Fig. \ref{fig:cutoffs} with a cutoff of 0 \AA, this gives the fitted temperature very close to the simulation temperature and the Maxwell distribution of velocities with $\alpha=2$.

\begin{figure}
{\scalebox{0.4}{\includegraphics{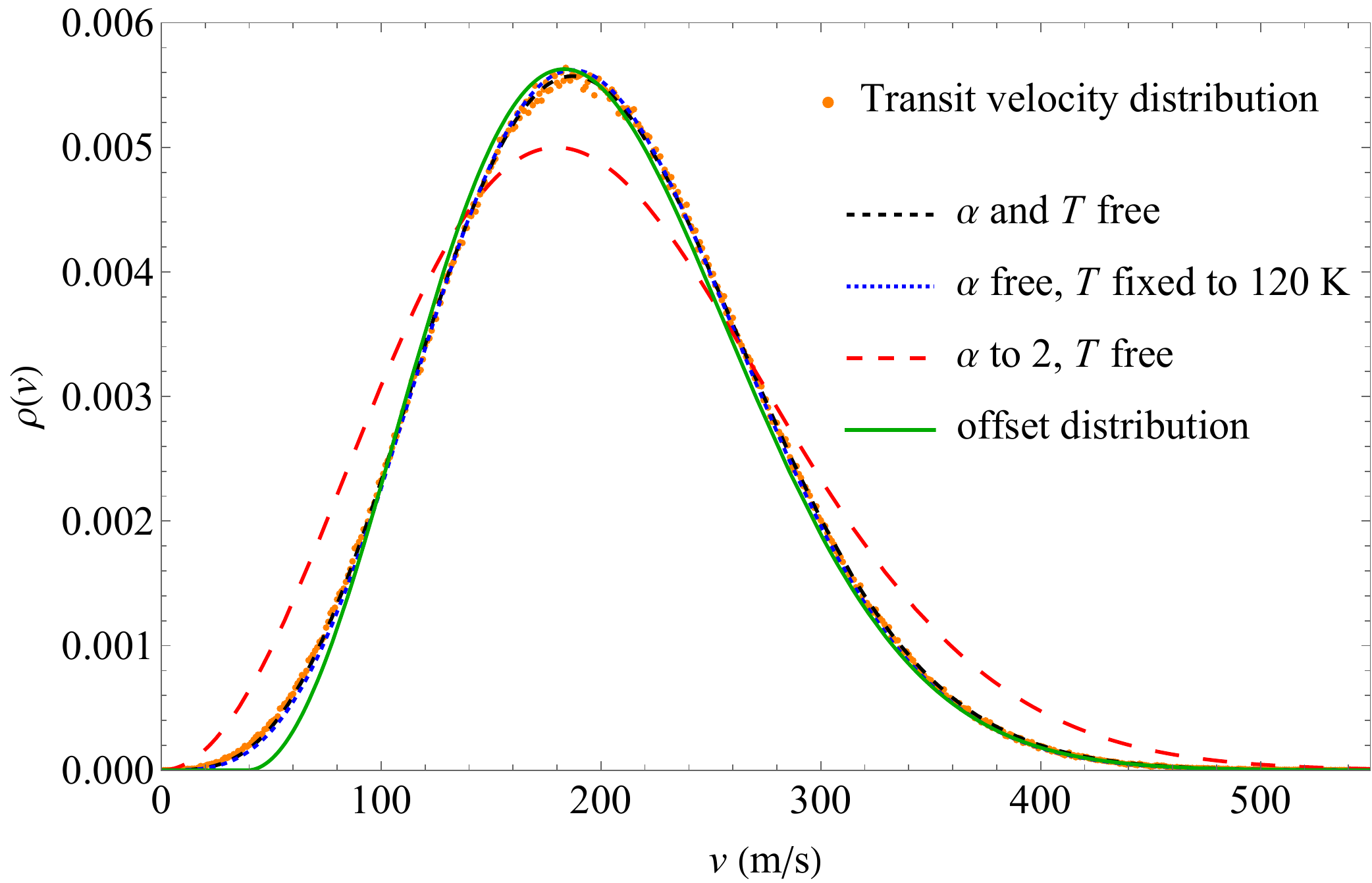}}}
\caption{Generalised Maxwell distributions from Eqs. \eqref{d1} and \eqref{eqn:offset} fit to the velocity distribution of transiting atoms at 100 bar and 25 K.}
\label{fig:distributioncompare}
\end{figure}

As seen in Fig. \ref{fig:distributioncompare}, $\rho(v)$ of transits can not be described by Eq. \eqref{d1} with $\alpha=2$: the fit is very poor. The fit markedly improves if we let $\alpha$ vary. This includes both cases where we fix $T$ to the simulation temperature and where we let $T$ vary. In the latter case, the fit quality is slightly better (this is not visible in Fig. \ref{fig:distributioncompare}) and returns temperature within 4\% agreement with the simulation temperature. At conditions in Fig. \ref{fig:distributioncompare}, $\alpha$ is about 3.25. 

According to Fig. \ref{fig:cutoffs}b, transits move on average faster than atoms in the overall system. The maximum of distribution \eqref{d1} corresponds to velocity $v_m$

\begin{equation}
v_m^2=\alpha\frac{k_{\rm B}T}{m}
\label{d2}
\end{equation}

Larger $v_m$ correspond to larger $\alpha$, consistent with our earlier funding of $\alpha=$3.25. 

Other modified Maxwell distributions were also employed in order to fit to the transit velocity distributions. To potentially describe the shift of velocities of faster transits to the right of the distribution, we fixed $\alpha$ and $T$, and offset the distribution by a fitting parameter:

\begin{equation}
    \label{eqn:offset}
    \rho(v) \propto \Theta(v-\mu) \ (v-\mu)^2 \exp\left( -\frac{m (v-\mu)^2}{2 k_{\mathrm{B}} T} \right)
\end{equation}

\noindent where $\mu$ is the offset parameter and the Heaviside step function $\Theta$ deletes the unphysical behaviour of the function outside the domain $v < \mu$. This resulted in a better fit, however the low-velocity tail of the distribution still does not match that of the transit velocity distribution. 


The above analysis indicates two important points. First, $\rho(v)$ of transits are not the same shape as that of a standard Maxwell distribution. This may not be unexpected in itself, in view that the detection algorithm imposes a condition on atoms to qualify as transits. What is notable, however, is that $\rho(v)$ are remarkably close to a generalised Maxwell distribution with $\alpha > 2$.

Second, we find that transits are on average faster than average motion of the system and therefore represent a distinct dynamical sub-ensemble of the overall system. This is seen in Fig. \ref{fig:cutoffs}b and is consistent with $\alpha>2$. In this picture, a liquid is a mixture of two distinct dynamical ensembles with different velocity distributions, albeit these ensembles are not fixed but dynamically change: an atom performs an oscillatory motion for time $\tau$ ($\tau$ is liquid relaxation time \cite{frenkel,dyre}), then becomes a transit atom, moving to a new position where it oscillates again and so on. These two distinct ensembles correspond to faster transit atoms and slower oscillating atoms, respectively. We note that this distinction between two sub-ensembles is lost if an average property over the entire system is calculated as is usually the case \cite{hansen1,balucani}.

\begin{figure}
{\scalebox{0.375}{\includegraphics{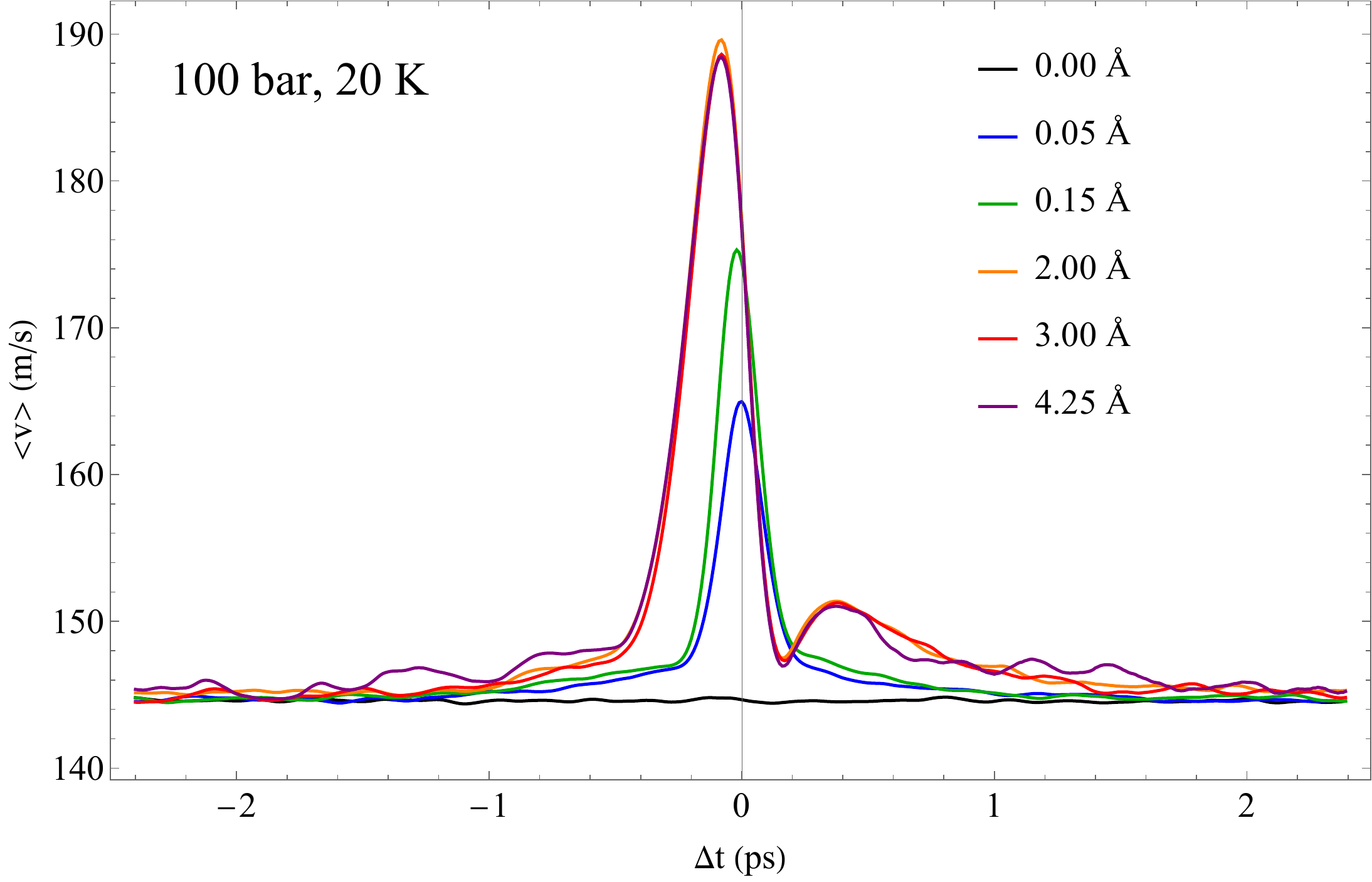}}}
\caption{Mean speed profile of transiting atoms for a window before and after the moment of transit as a function of transit cutoff definition.}
\label{fig:vprofile}
\end{figure}

We characterise the nature of a transit by examining the speed of atoms in a time window around the moment of transit. This is done by collecting velocity data from atoms in times preceding and following the transit as per our above definition. At long times before and after the transit, the mean speed is expected to return to the mean speed defined by the Maxwell distribution, as transiting is expected to happen to all atoms with equal probability, such that individual atoms are not more prone to transiting than other atoms. 
We plot the speed profile averaged over 200,000 jumping atoms at 100 bar and 20 K, just above the melting line, in Fig. \ref{fig:vprofile}. This profile extends 2.4 ps before and after the moment of transit, defined by the above condition, which defines $\Delta t = 0$. This is done for several transit cutoffs, demonstrating the same cutoff-independence seen in Fig. \ref{fig:cutoffs}. We observe that the mean speed is peaked just before the moment of transit, implying that atoms accelerate out of their local environment into a new environment, facilitated by the opening up of a fast diffusion pathway. The subsequent minimum and secondary maximum after this peak represents the deceleration as the atom settles into its new oscillatory quasi-equilibrium position. At long times, the mean speed becomes identical to the mean speed of the Maxwell distribution, seen in the flat line of the 0 \AA\ cutoff. Small but nonzero cutoffs deviate from this velocity and converge on the invariant speed profile at larger cutoffs, as was seen in distributions in Fig. \ref{fig:cutoffs}. We note that this speed profile is not time-symmetric: the minimum and secondary maximum are not seen \textit{before} the transit time, and the peak is not centred on $\Delta t = 0$. We also note that the peak in Fig. \ref{fig:vprofile} is seen for small as well as large cutofffs. Recall that the transit is identified by its displacements from the starting position (in the transiting atom's past, which is the source of the asymmetry). Hence both fast and slower atoms are counted as transits for short cutoffs, resulting in smaller average peak velocity as compared to larger peak related to long cutoffs.


These separate ensembles are a contrast to the motion in solids and gases. In solids, all motion is oscillatory and no separate regime can be defined. As a result, the velocity distribution is Maxwellian. In gases where collisions are rare and effectively instantaneous, atoms spend most of their time moving freely with velocities similarly distributed according to the Maxwell distribution.

We add three remarks regarding $\rho(v)$ of transits. First, there is no apriori clear reason to assume that transits in liquids are on average faster. One general consideration might be as follows. No jumps can take place in any reasonable time in a liquid at fixed cage volume because this would involve large energy barriers and long waiting times. Instead, Frenkel argued \cite{frenkel}, a volume increase of a surrounding fluctuating cage provides a fast diffusion pathway for an atom to escape and become a transit. Traversal of these diffusion pathways is associated with a higher kinetic energy due to activation energy barriers involved. If we assume the speed of an oscillating atom takes the form $v = v_0 \sin(\omega t)$, then the average speed is $\frac{2}{\pi} v_0$ over the oscillation period. If the transit speed, on the other hand, is comparable to the amplitude, $v_0$ transits are expected to be on average faster than oscillating atoms by a factor of $\frac{\pi}{2}$. Whereas this is clearly an approximate argument, the direct quantitative evidence for faster transits comes from our current MD simulations.


Second, the non-Maxwellian $\rho(v)$ of transits does not contradict the Maxwell distribution in the entire classical system originating from the factorisation of distributions dependent on momenta and coordinates, respectively \cite{landaustat}. As mentioned earlier, $\rho(v)$ of transits corresponds to a dynamically changing sub-ensemble of the whole system which has the Maxwell distribution overall.

Third, the ratio of transits identified by the algorithm to the overall number of atoms is small: for example, at 40 K and 1 kbar, just above the melting line, about 1\% of atoms within a given 0.1 ps window, approximately corresponding to a typical Debye vibration period, are in-transit, meaning that over the course of an oscillation an atom has a 1\% probability to be a transit in the detection algorithm. In the gaslike state at 200 K and 1 kbar, this increases to 8\%. For this reason, the distribution of atoms tagged as non-transits is very close to the velocity distribution in the overall system which is Maxwellian.

\begin{figure}
{\scalebox{0.55}{\includegraphics{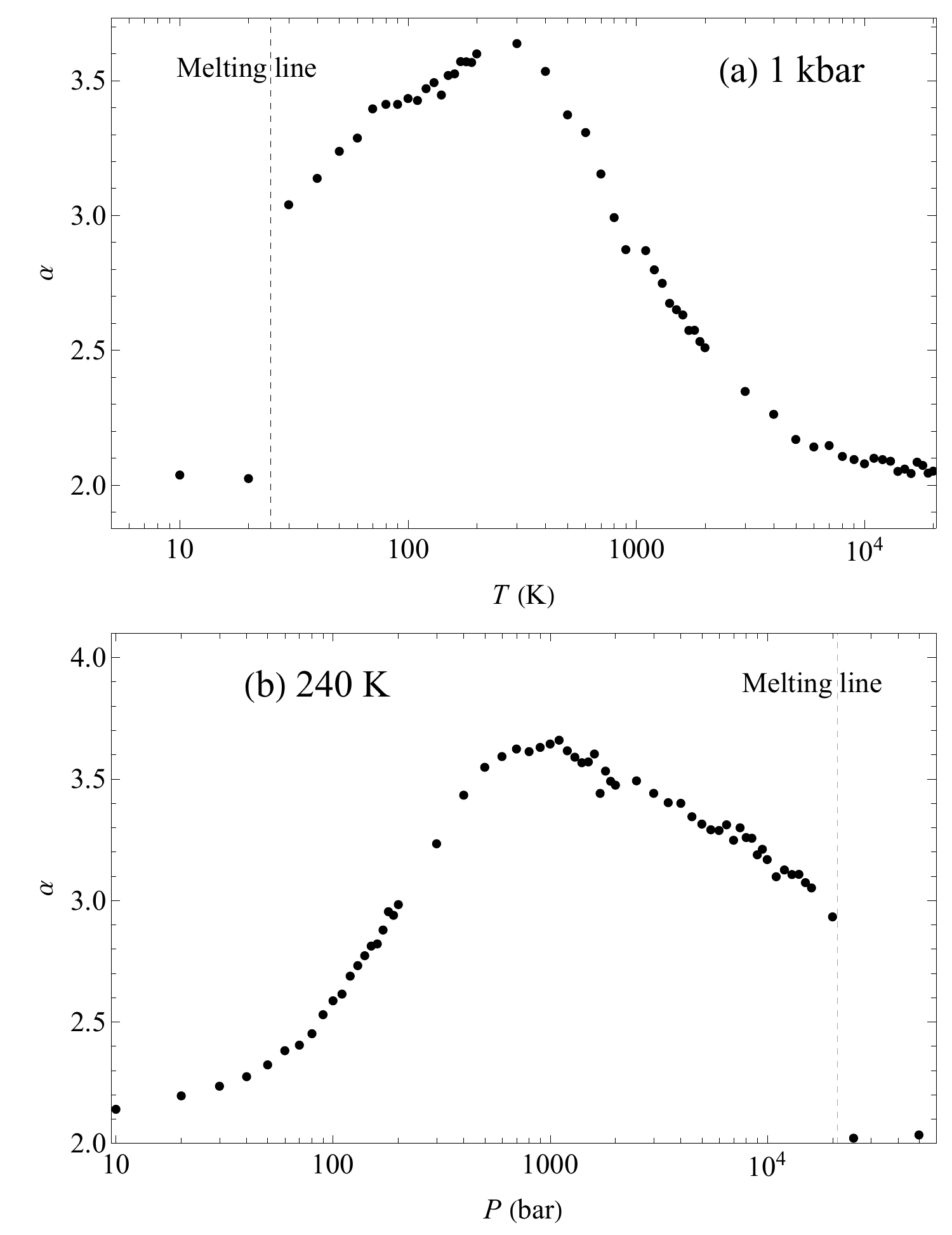}}}
\caption{Parameter $\alpha$ in the generalised Maxwell distribution as a function of: (a) temperature; (b) pressure on the 1 kbar isobar and 240 K isotherm respectively.}
\label{fig:alpha}
\end{figure}

We now address the variation of $\alpha$ with temperature and pressure. In Fig. \ref{fig:alpha}, we plot $\alpha$ as a function of temperature varied in a wide range including both the subcritical liquid and the supercritical fluid state. We observe an interesting non-monotonic behavior with a maximum. At low temperature where the system is a solid, our algorithm does not detect transits, and therefore the effective dimensionality is undefined. The calculated distribution of velocities over all atoms in the solid phase is Maxwellian with $\alpha=2$ as expected, which is what is shown below the melting line in Fig. \ref{fig:alpha} for reference. 

At high temperature in the gaslike state, $\alpha$ tends to $2$ as well. This takes place on both isobars and isotherms. At these extreme conditions above the Frenkel line \cite{flreview}, an atom lacks any quasi-equilibrium environment, corresponding to the complete loss of all minima in the RDF, as seen in Fig. \ref{fig:cutoffs}, with only a very broad and low first peak remaining. In this high-temperature/low-pressure gaslike regime, particle dynamics is governed by the gaslike ballistic motion where collisions set the particle mean free path \cite{chapman}. Velocity distribution is Maxwellian in this case, causing the return of $\alpha$ to 2. 

The value of $\alpha$ in Eq. \eqref{d1} is related to space dimensionality: the Maxwell distribution of velocity components $v_x$, $v_y$ and $v_z$ in three-dimensional space corresponds to $\rho(v)\propto v^2$ in spherical coordinates \cite{landaustat}. A Maxwell distribution in two spatial dimensions has the exponent $\alpha = 1$, and in four dimensions $\alpha = 3$. Therefore, $\alpha>2$ corresponds to the motion in effectively higher dimensionality. 

This higher effective dimensionality is related to the additional condition imposed upon a transit. This condition is associated with the higher energy involved in the particle escape from the surrounding cage: the velocity distribution of transits is deformed towards faster $v$ as is seen in Fig. \ref{fig:cutoffs}b. We have seen that this deformation can not be represented by a higher temperature, but only by a higher space dimensionality. $\alpha=2$ in the Maxwell distribution in Eq. \eqref{d1} corresponds to a geometric factor $v^2$, arising from the two-dimensional surface corresponding to a single speed in a uniform three-dimensional velocity space. Deformed towards faster $v$, the speed distribution of transits can also be represented in a uniform velocity space, albeit with higher dimensionality. 

As is seen in Fig. \ref{fig:alpha}, $\alpha$ is generally not an integer but is fractional. The largest $\alpha$ is close to $4$ ($\alpha=3.7$) and corresponds to an effective space dimensionality close to five. In view that $\alpha$ is not generally an integer, it is interesting to note that close and above the melting line along the isobar and isotherm in Fig. \ref{fig:alpha}, $\alpha$ is close to 3, such that the subsystem of transits follows a Maxwell distribution in a four-dimensional space. $\alpha=3$ at melting could be a potentially useful insight for a theory of the solid-liquid melting transition, in view that mobile transiting defects aid the formation of the liquid nucleus and lower the barrier for melting \cite{tuckerman}. 

In a general sense, the existence of two dynamical sub-ensembles might resemble dynamic heterogeneity (DH) noted in slow glassy dynamics of viscous supercooled liquids approaching glass transition \cite{ediger}. This DH can originate from the expected correlation of mobility and structure: higher mobility is caused by fewer neighbours and wider surrounding cage due to the nature of the model used, which was heterogeneous with two different atomic species \cite{hetero1,hetero2}. This is a model-specific and hence a non-generic effect. The effect reported here is unrelated to DH in glassy dynamics and operates in a markedly different part of the phase diagram. For example, the maximum of $\alpha$, corresponding to fast dynamics where the deviation from the Maxwell distribution is the largest, takes place well above the critical point, namely at $6.8T_c$ in Fig. \ref{fig:alpha}a and $4.5P_c$ in Fig. \ref{fig:alpha}b, where $T_c$ and $P_c$ are critical temperature and pressure, 45 K and 27 bar respectively in this system. This effective dimensionality is therefore a universal effect, operating both in subcritical liquids and deep in the supercritical state, rather than solely in supercooled liquids. Differently from effective higher dimensionality reported here, DH in supercooled liquids due to structural restraints can be anisotropic and lower-dimensional \cite{harrowell2022}. The velocity profile of transiting atoms, presented in Fig. \ref{fig:vprofile}, is calculated over transiting atoms, with no reference to their local environments, and shows the essence of these transits: a sharp increase in speed during the transit with rapid decorrelation to the average thermal speed on either side. Indeed, transiting atoms are not spatially correlated with each other. Every atom in the fluid will undergo a transit, given enough time, and the average speed of this atom during the transit will obey the profile displayed in Fig. \ref{fig:vprofile}.

We note that topological quantum matter in $D$ real spatial dimensions can be described by effective high dimensionality $D+d$, where $d$ are synthetic dimensions related to additional distinct degrees of freedom due to, for example, coupling, and arise through reinterpretation of underlying physics \cite{naturereview,phystoday}. This is generally similar to our current finding: higher effective dimensionality in liquids is related to a distinct (faster) dynamical sub-ensemble in the system. Differently to higher synthetic dimensions discussed earlier \cite{naturereview}, higher dimensionality in liquids does not need to be integer and is fractional. 

\section{Summary}

In summary, we have shown that flow-enabling transit atoms in liquids form a dynamically distinct sub-ensemble where atoms move on average faster than the overall system and which has a manifestly non-Maxwellian velocity distribution. This is in contrast to solids and gases where no such distinction related to different ensembles can be made. The non-Maxwellian distribution of transits can be described by an exponent $\alpha$ corresponding to high space dimensionality. This dimenstionality is close to 4 at melting and exceeds 4 at high temperature. $\alpha$ in the liquid state interestingly has a maximum as a function of temperature and pressure, returning to its Maxwell value in the solid and gas states.

\section{Data Availability}
The authors declare that the data supporting the findings of this study are
available within the paper. 

\section{Acknowledgements}

We thank Peter Harrowell for critical reading of the manuscript and for useful comments.

\section{Author Contributions}
All authors conceptualised the study. C.C. performed simulations and subsequent data analysis. All authors designed and wrote the manuscript.

\section{Competing Interests}
The authors declare no competing interests.

\bibliography{refs}

\begin{thebibliography}{10}

\bibitem{landaustat}
L.~D. Landau and E.~M. Lifshitz.
\newblock {\em Statistical Physics, part 1.}
\newblock Pergamon Press, 1970.

\bibitem{Pitaevskii}
L.~P. Pitaevskii.
\newblock Statistical physics. liquids.
\newblock {\em Physical Encyclopedia, ed. by A. M. Prokhorov}, 4:670, 1994.

\bibitem{akhiezer}
A.~I. Akhiezer.
\newblock Memories about jacob ilyich frenkel.
\newblock {\em Low Temp. Phys.}, 20:194, 1994.

\bibitem{ropp}
K.~Trachenko and V.~V. Brazhkin.
\newblock Collective modes and thermodynamics of the liquid state.
\newblock {\em Reports on Progress in Physics}, 79:016502, 2016.

\bibitem{frenkel}
J.~Frenkel.
\newblock {\em Kinetic Theory of Liquids}.
\newblock Oxford University Press, 1947.

\bibitem{wallacemd}
D.~C. Wallace, E.~D. Chisolm, and B.~E. Clements.
\newblock Observation of single transits in supercooled monatomic liquids.
\newblock {\em Phys. Rev. E}, 64:011205, 2001.

\bibitem{dyre}
J.~C. Dyre.
\newblock Colloquium: The glass transition and elastic models of glass-forming
  liquids.
\newblock {\em Rev. Mod. Phys.}, 78:953, 2006.

\bibitem{hansen1}
J.~P. Hansen and I.~R. McDonald.
\newblock {\em Theory of Simple Liquids}.
\newblock Elsevier, 2013.

\bibitem{balucani}
U.~Balucani and M.~Zoppi.
\newblock {\em Dynamics of the Liquid State}.
\newblock Oxford University Press, 2003.

\bibitem{copley}
J.~R.~D. Copley and J.~M. Rowe.
\newblock Short-wavelength collective excitations in liquid rubidium observed
  by coherent neutron scattering.
\newblock {\em Phys. Rev. Lett.}, 32:49, 1974.

\bibitem{pilgrim}
W.~C. Pilgrim, S.~Hosokawa, H.~Saggau, H.~Sinn, and E.~Burkel.
\newblock Temperature dependence of collective modes in liquid sodium.
\newblock {\em J. Non-Cryst. Sol.}, 250-252:96, 1999.

\bibitem{burkel}
E.~Burkel.
\newblock Phonon spectroscopy by inelastic x-ray scattering.
\newblock {\em Rep. Prog. Phys.}, 63:171, 2000.

\bibitem{pilgrim2}
W.~C. Pilgrim and C.~Morkel.
\newblock State dependent particle dynamics in liquid alkali metals.
\newblock {\em J. Phys.: Condens. Matt.}, 18:R585, 2006.

\bibitem{water-tran}
A.~Cunsolo, C.~N. Kodituwakku, F.~Bencivenga, M.~Frontzek, B.~M. Leu, and A.~H.
  Said.
\newblock Transverse dynamics of water across the melting point: a parallel
  neutron and x-ray inelastic scattering study.
\newblock {\em Phys. Rev. B}, 85:174305, 2012.

\bibitem{hoso}
S.~Hosokawa et~al.
\newblock Transverse acoustic excitations in liquid ga.
\newblock {\em Phys. Rev. Lett.}, 102:105502, 2009.

\bibitem{hoso3}
S.~Hosokawa, M.~Inui, Y.~Kajihara, S.~Tsutsui, and A.~Q.~R. Baron.
\newblock Transverse excitations in liquid sn.
\newblock {\em J. Phys.: Condens. Matt.}, 27:194104, 2015.

\bibitem{monaco1}
V.~M. Giordano and G.~Monaco.
\newblock Fingerprints of order and disorder on the high-frequency dynamics of
  liquids.
\newblock {\em PNAS}, 107:21985, 2010.

\bibitem{monaco2}
V.~M. Giordano and G.~Monaco.
\newblock Inelastic x-ray scattering study of liquid ga: Implications for the
  short-range order.
\newblock {\em Phys. Rev. B}, 84:052201, 2011.

\bibitem{sn}
S.~Hosokawa et~al.
\newblock Transverse excitations in liquid sn.
\newblock {\em J. Phys.: Condens. Matt.}, 25:112101, 2013.

\bibitem{chapman}
Sydney Chapman and T.~G. Cowling.
\newblock {\em {The mathematical theory of non-uniform gases}}.
\newblock Cambridge University Press, 1990.

\bibitem{naturereview}
T.~Ozawa and H.~M. Price.
\newblock Topological quantum matter in synthetic dimensions.
\newblock {\em Nature Rev. Phys.}, 1:349, 2019.

\bibitem{phystoday}
K.~R.~A. Hazzard and B.~Gadway.
\newblock Synthetic dimensions.
\newblock {\em Physics Today}, 76(4):62, 2023.

\bibitem{dlpoly}
I.~T. Todorov, B.~Smith, M.~T. Dove, and K.~Trachenko.
\newblock {\em J. Mater. Chem.}, 16:1911, 2006.

\bibitem{sastry}
S.~Sengupta, F.~Vasconcelos, F.~Affouard, and S.~Sastry.
\newblock Dependence of the fragility of a glass former on the softness of
  interparticle interactions.
\newblock {\em The Journal of Chemical Physics}, 135:194503, 2011.

\bibitem{egami}
T.~Iwashita, D.~M. Nicholson, and T.~Egami.
\newblock Elementary excitations and crossover phenomenon in liquids.
\newblock {\em Physical Review Letters}, 110:205504, 2013.

\bibitem{flreview}
C.~Cockrell, V.~V. Brazhkin, and K.~Trachenko.
\newblock Transition in the supercritical state of matter: review of
  experimental evidence.
\newblock {\em Physics Reports}, 941:1, 2021.

\bibitem{tuckerman}
A.~Samanta, M.~E. Tuckerman, T.~Q. Yu, and E.~Weinan.
\newblock Microscopic mechanisms of equilibrium melting of a solid.
\newblock {\em Science}, 346:729, 2014.

\bibitem{ediger}
M.~D. Ediger.
\newblock Spatially heteregogeneous dynamics in supercooled liquids.
\newblock {\em Annu. Rev. Phys. Chem.}, 51:99, 2000.

\bibitem{hetero1}
K.~Vollmayr-Lee, W.~Kob, K.~Binder, and A.~Zippelius.
\newblock Dynamical heterogeneities below the glass transition.
\newblock {\em J. Chem. Phys.}, 116:5158, 2002.

\bibitem{hetero2}
M.~S.~G. Razul, G.~S. Matharoo, and P.~H. Poole.
\newblock Spatial correlation of the dynamic propensity of a glass-forming
  liquid.
\newblock {\em J. Phys.: Condens. Matt.}, 23:235103, 2011.

\bibitem{harrowell2022}
G.~Sun and P.~Harrowell.
\newblock A general structural order parameter for the amorphous solidification
  of a supercooled liquid.
\newblock {\em J. Chem. Phys.}, 157:024501, 2022.

\end{thebibliography}

\bibliographystyle{unsrt}

\end{document}